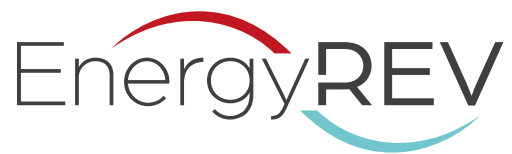

# The energy revolution: cyber physical advances and opportunities for smart local energy systems

Nandor Verba, Elena Gaura, Stephen McArthur,
George Konstantopoulos, Jianzhoug Wu,
Zhong Fan, Dimitrios Athanasiadis,
Pablo Rodolfo Baldivieso Monasterios,
Euan Morris and Jeffrey Hardy

June 2020

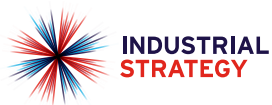
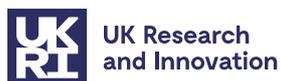

# Acknowledgements

Nandor Verba | Centre for Data Science, Coventry University, UK

Elena Gaura | Centre for Data Science, Coventry University, UK

Stephen McArthur | Department of Electronic and Electrical Engineering, University of Strathclyde, UK

George Konstantopoulos | Department of Automatic Control & Systems Engineering, University of Sheffield, UK

Jianzhoug Wu | Department of Electrical and Electronic Engineering, Cardiff University, UK

Zhong Fan | School of Computing and Mathematics, Keele University, UK

Dimitrios Athanasiadis | School of Computing and Mathematics, Keele University, UK

Pablo Rodolfo Baldivieso Monasterios | Department of Automatic Control & Systems Engineering, University of Sheffield, UK

Euan Morris | Department of Electronic and Electrical Engineering, University of Strathclyde, UK

Jeffrey Hardy | Grantham Institute, Imperial College London, UK

This report should be referenced as:

Verba, N., Gaura, E., McArthur, S., Konstantopoulos, G., Wu, J., Fan, Z., Athanasiadis, D., Monasterios, P.R.B., Morris, E. and Hardy, J. 2020. The energy revolution: cyber physical advances and opportunities for smart local energy systems. EnergyREV, University of Strathclyde Publishing: Glasgow, UK.
ISBN 978-1-909522-58-9

# Contents

# Future–proofing cyber physical architectures

## The context and need

As we recognise the need to transform our future energy systems, there is a rapid acceleration in concept designs, demonstrator systems and early-adopter installations. These all attempt to harness low carbon energy generation and dynamically deal with the challenges around generation, transmission, distribution, and end use from an increasingly whole-system and multi-vector viewpoint.

The energy systems are often developed for a specific range of use cases and functions, and these match the requirements and needs of the community, location or site under consideration. During the design and commissioning, new and dedicated cyber physical architectures are developed. These are the control and data systems that are needed to bridge the gap between the physical assets (e.g. generation technologies, electrical equipment, heat equipment, electricity and heat infrastructure, etc.) the data captured, and the control signals sent. Often, the cyber physical architecture and infrastructure is focused on functionality and the delivery of the detailed technical specifications without consideration of future adaptations.

Taking a longer-term view, our assertion is that the cyber physical architecture can be developed such that it provides three further key capabilities:

- **Flexibility**: Throughout their lifecycle new energy systems, and especially smart local energy systems, will be subject to: a growing set of use cases; connection of new devices and assets; new data systems/data analysis requirements; increased end-users (with new information needs or who interact with the energy system in a new way) or additional energy vectors being considered. It is important to ensure that energy systems are flexible so that they can accommodate these changes without a re-engineering of the cyber system.
- **Scalability**: The energy system will need to deal with physical extensions through additional generation technologies or energy delivery devices. They may also be extended through additional energy vectors to be taken on board.
- **Reusability**: The overall system may need to be transplanted to similar problems in other locations.

There are technologies and approaches that have arisen from other fields that, if used within SLES, could support flexibility, scalability and interoperability. As these can improve the operational data systems then they can also be used to enhance predictive functions. Alongside this, the new methods can support resilience and fault tolerance, which is essential for a secure energy system. All of the functionality needs to draw upon the state-of-the-art in data privacy and data security. If used and deployed effectively, these new approaches can offer longer term improvements in the use and effectiveness of SLES, while allowing the concepts and designs to be capitalised upon through wider roll-out and the offering of commercial services or products.

Therefore, our view is that it is vital to consider how to make energy systems flexible, scalable, interoperable, predictive and secure. In order to do this, improved cyber physical architectures are essential.

## Our approach

We have designed a two stage, 10 step process to give organisations a method to analyse SLES projects based on their Cyber Physical System (CPS) components and develop a future-proof energy system.

### A. Functional Analysis Framework

1. Set out the background and context for the SLES design
2. Identify targeted areas of improvement against the current industry practice
3. Identify proposed benefit generated by SLES against the industry norm
4. Produce a system architecture overview
5. Identify the smart properties
6. Review the technical and technological requirements of implementing the SLES design
7. Develop a key performance indicator Key Performance Indicator (KPI) map

### B. Expansion Framework

1. Determine the future technical and technological requirements
2. Develop the new system architecture
3. Evaluate the extended architecture and functionality improvements

This type of analysis allows comparisons to be made between SLES, positioning individual SLES against academic and industrial state of the art and ultimately identifying and prioritising benefit adding components and artefacts that can be retrofitted to extend a given system.

An evaluation of the **ADvanced multi-Energy management and oPTimisation time shifting platform** (ADEPT) shows how the process can be used to evaluate and extend an existing SLES.

A sound understanding of the potential of advanced CPS technologies, together with the simple framework for SLES evaluation during the design process, can lead to improved system performance. This ensures that SLES can withstand the test of time.

# A cyber–physical system analysis framework for energy systems

This demonstrator design analysis framework aims to provide a systematic method of evaluating energy projects based on key parameters and considerations that researchers focusing on Cyber Physical Systems and SLES have identified. The framework highlights key architecture and technology traits that are crucial for organizations to have transparency over when seeking to understand, adapt and grow energy systems.

This framework is designed in much the same way as a Strengths, Weaknesses, Opportunities, and Threats (SWOT) analysis. It should be completed at different stages of the digital system architecture development, deployment and use of the energy system. The process we devised and associated framework can be re-applied to system designs when new technologies emerge in order to identify when upgrades and extensions may be required or beneficial.

The smart attributes of a SLES are usually linked to its CPS components; these entail all the devices and sub-systems that are designed to digitise the local energy system, make it more connected and allow a better understanding of its functional condition and performance. As outlined earlier, the CPS components need to deliver: flexibility; scalability; interoperability; resilience and fault tolerance; prediction; privacy and security.

Such functional traits are delivered through architectural components and technologies drawn from fields such as AI, distributed control, cyber security, sensing and state estimation, advanced computing and data transport. Our framework is designed to identify where the advances in these fields can or need to be integrated into a SLES. The gaps and opportunities in each area can highlight where existing innovations can bring impact and the likely benefits that will arise.

# A. The functional Analysis Framework explained

Fig 1 shows a high-level view of the Analysis framework. It has 7 sections that are designed to bring to the fore the key digital elements and properties of demonstrators that are required to future proof and/or extend the capabilities of the SLES under consideration. Exploring the framework in greater detail reveals the following sections, or steps to be undertaken when evaluating SLES with a look to the future.

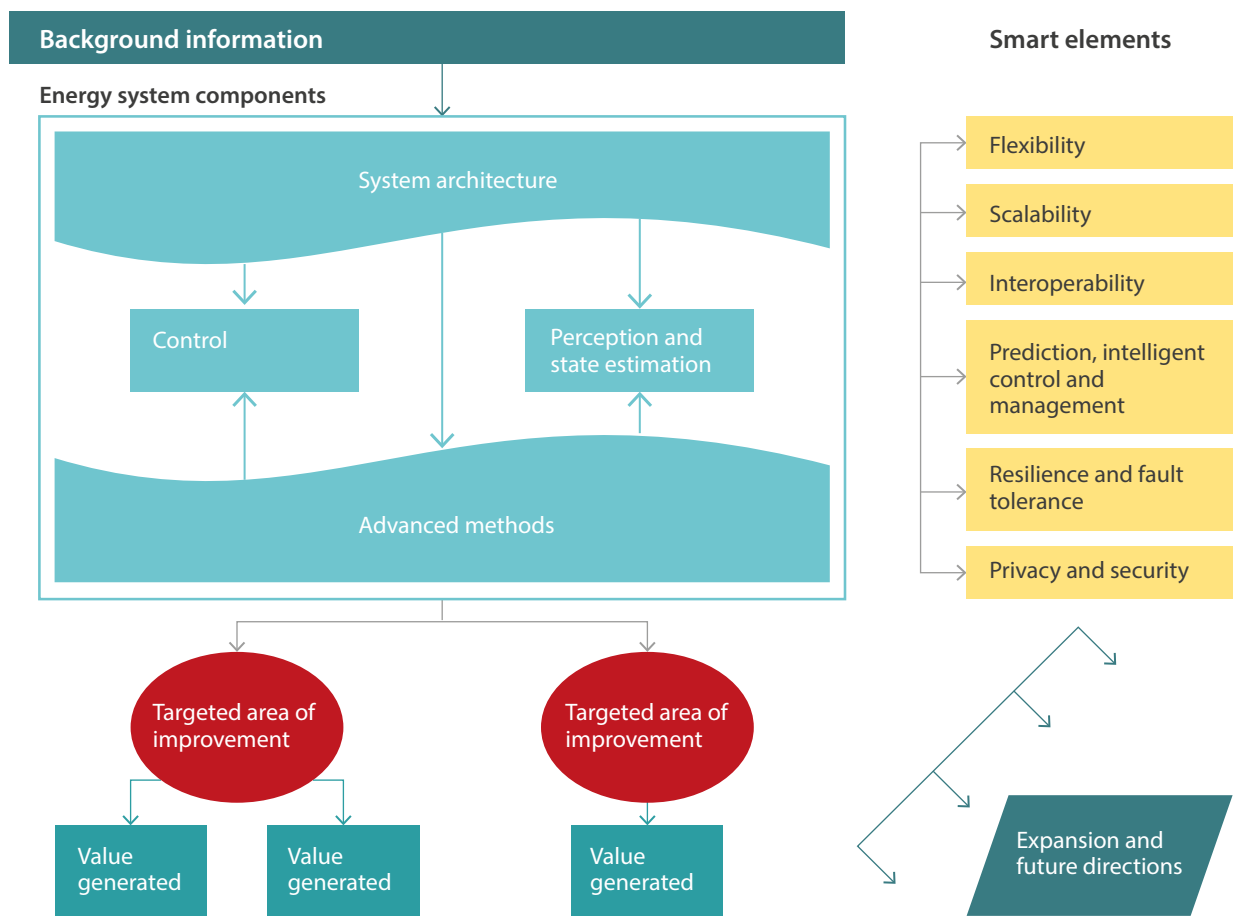

Figure 1: High-level view of Analysis framework.

## 1 Setting out the background and context for the SLES design

This step sets out the key elements of the demonstrator, including partners, funding, high-level objectives and timeline for implementation as well as basic considerations for the design. While this is commonly known data within a team, laying this out as a first step in the evaluation process is necessary to align all stakeholders with the vision for the design.

## 2 Identify targeted areas of improvement against the current industry practice

This step analyses the motivation for the energy demonstrator design and indicates the state-of-the-art in the application area. The key task is to break down the architecture into its energy system technical components (as shown in the Framework in Fig 1) and clearly define the functional advance or improvement that each one is being designed to deliver.

## 3 Identify proposed benefit generated by SLES against the industry norm

This step will identify the novel aspects of the demonstrator and show how these enhance the operation of the SLES. The outcomes should also highlight how key performance indicators (KPIs) of SLES were met. These KPIs could range from Manufacturing Readiness Level (MRL) increments, to increased features and showcase attributes, to technical KPIs such as reducing the energy loss in a system by a specified amount.

## 4 Produce a system architecture overview

The system architecture step aims to produce a high-level diagrammatic representation of the system, that contains stakeholders' agreed views on information related to the architectural components, how they are interconnected and any other relevant information that the team can use to take this system further in terms of future proofing and extending its capabilities. The diagrams can contain the integration of and interaction between energy subsystems in both the physical (e.g. energy infrastructure) and the cyber (e.g. information exchange, communication links, etc) layers. Data life-cycle reviews, network diagrams, data and communication pipelines, sensor and state maps, and so on may be considered in this step.

## 5 Identify the *smart* properties

This step defines how certain properties and desired features of SLES are (or are to be) implemented in the demonstrator.

For all of these desired features, there is the opportunity to take advantage of recent advances in ICT, data management systems and middleware, AI algorithms, distributed AI, and distributed control to move energy systems to the next level of smart functionality.

### Flexibility

Flexibility, in the context of this framework, is the characteristic of a system architecture or process that allows it to be re-configured and adapted for various use-cases and deployment scenarios. It allows components to be easily modified and changed without the need to replace the whole setup.

The use of loosely coupled digital elements and processes is a major step towards creating flexible environments. Breaking large cyber physical systems down into components and services allows whole processing elements and functions to be migrated and replaced with minimal disruption.

Throughout their lifecycle SLES are expected to accommodate: new devices and protocols; digital and physical extensions; changes in the system size and the variety of energy vectors; changes in sources and consumers.

## Scalability

The scalability properties of SLES are those that allow architectures to cope with an increased number of connected devices and resources. The devices can vary from energy generation units to storage and transformation elements, to monitoring, control and communication elements. The scaling of resources can vary between an increase in the amount of energy resources managed or monitored, to an increase in physical resources such as houses, cars, sensors and devices.

With an increase in physical resources there is also an increase in data gathering, orchestration and maintenance of data sources and data streams. The cyber-physical architectures behind these resources need to handle the increase in demand.

This requires a flexible and loosely coupled system architecture that can take advantage of novel orchestration approaches. A complete system can monitor loads and make decisions on where services are being deployed and where data is being stored. A hybrid of local and centralised solutions can offer the scalability that is required while maintaining robustness constraints.

## Interoperability

Interoperability ties in with the horizontal integration of businesses where products and components are developed in such a way that they can interact and complement products from other suppliers. A common approach for interoperability is the use of standard protocols and designs. This approach has a lot of merit. Even if two systems use varying standards, conversion and translation mechanisms are more likely to exist between two standard elements.

Industry protocols are fit for purpose, designed to be reliable and robust. They don't lend themselves to flexible and loose connections, as they are not designed for that. The use of brokers and gateways to bridge these networks to allow for more advanced computation and analytics elements to be used is the most common approach. These gateways can bridge industry standard protocols to ones commonly used in Big Data and AI applications. This may allow the internal network to retain its robustness and reliability while taking advantage of advanced computational elements.

## Prediction, intelligent control and management

Adding capability to an energy system to predict consumption and generation is a key element of providing a resilient and efficient supply. The smart element focuses on not just the algorithms and methods, but also on how these can be deployed and how a range of these can be tested. This element also looks at how functionality from various providers can be integrated to meet local needs. A fully functioning smart system needs to embed elements of power use and generation forecasting within market tools. There is also the need to intelligently balance or optimise control and management decisions around the use of energy sources, the energy demand, and any specific constraints on equipment and devices. In addition, these need to inform maintenance and asset management decisions. A range of new algorithms will need to be integrated into the energy systems to deliver this.

Bridging industry standard approaches in energy systems to commonly used protocols in ICT has the advantage of connecting the system to a wide range of available monitoring and analysis platforms to support this. These offer a set of tools, libraries or data flows that can allow the user to feed real-time data into their simulations, models, optimisations and algorithms. Storing the data in an appropriate manner and having a well-designed processing system on top allows for efficient and fast historical data processing.

### Resilience and fault tolerance

The capability of a SLES to recover and maintain full or reduced functionality even with the loss of digital or physical components is critical in attaining the reliability required to increase their adoption.

From the digital service perspective, there are several aspects that can be considered. For example, the flexibility built into the system may allow for graceful degradation if digital devices and processes fail. In addition, under certain circumstances, it may be possible to move functions and processes from local deployments to centralised ones in a non-disruptive manner. This requires local and centralised duplications of data and digital components but offers the ability to switch between deployments to suit local requirements or deal with resilience issues. The ability to move digital resources onto secondary or back-up processing elements and communication channels is also crucial in providing increased reliability and services availability.

### Privacy and security

Cyber security is a key component when taking a system design innovative concept to a public test bed and into production. The level of cyber security that a system needs can vary based on the sensitivity of the data and the capability of the underlying architecture.

The management of data throughout SLES systems is critical to ensure public confidence and system integrity. The security measures at the processing level of the architecture where the servers and data centres lie are generally robust and well understood by designers. At the edge of the network, where the peripheral devices and gateways lie, the situation may be different. These devices are more resource constrained and might not have firewalls on their networks.

## 6 Review the technical and technological requirements of implementing the SLES design

This technical step aims to bring to the fore the technologies needed to implement the existing system. The review helps to highlight and acknowledge the Technology Readiness Level (TRL )of these technologies and allows stakeholders to view the demonstrator from a business-oriented perspective. It further highlights the limitations and advantages of the employed technologies.

## 7 Develop a KPI Map

The KPI map is a combination of the information and outcomes from the previous sections, that boils these down to an industry friendly graph that connects roadmap and high-level directional elements with measurable KPI's, technologies and research directions.

# B. The Expansion Framework explained

There are 3 basic steps to the Expansion Framework:

## 1 Determine the future technical and technological requirements

This step highlights the need for the expansion. It can be used to show what new or existing requirements the demonstrator failed to achieve or couldn't fully achieve. It looks at the limiting factors in the project, what new elements can be added and what features could be improved upon. More importantly, it also allows for reflection on the benefit these improvements would generate. The aim here is to allow researchers and industry to get a glimpse of niches and areas where development is required.

## 2 Develop the new system architecture

This step is designed to provide a detailed design of the expansion. It should contain detailed and high-level diagrammatic representation of the system pertaining to stakeholders, technologies and any information that would be required to implement such a system or to understand its benefits and direction. Its diagrams and contents can be formulated in the same way as the system architecture overview in step 4 of the Analysis framework above. Highlights of the TR and MR levels of each component and proposed innovation could be noted and acknowledged, together with risks involved in the expansion.

## 3 Evaluate the extended architecture and functionality improvements

This step is a reflective element that aims to draw conclusions for the expansion work. It can be used to highlight shortcomings of the new designs or areas where the technology has not yet reached the required maturity or integration readiness. The outputs of this step can lead to re-drawing the KPI map to reflect the extensions or can serve as a place to show all the benefits and added functionality an extension can offer.

# Putting the framework to work: the ADEPT Demonstrator Analysis

The ADEPT Demonstrator use-case analysis shows how the framework should be applied and how it can be used to extend and evaluate a system. In this case the evaluation will be used by external members of the original design team to better understand the demonstrator and be able to extend the cyber-physical elements of ADEPT in order to add extra layers of digitalisation. Specifically, this is to facilitate the deployment of Distributed Controllers, Dashboarding, Storage and MAS systems.

## 1 Setting out the background and context for the SLES design

The main aim of the Advanced multi-Energy management and oPTimisation time shifting platform (ADEPT) is to create an intelligent micro-grid system based on several Distributed Energy Resources (DERs) that combine a variety of smart features. It is one of a few SLES that have been designed to solve the problems caused by the uncoordinated increase of energy source integration without sophisticated operation mechanisms.

ADEPT provide a solution to these problems by using clean and affordable energy within a local community of energy producers and consumers in an optimal manner to benefit every community member. Within this framework, several local energy demonstrators have been developed in the past years to combine multiple energy resources, e.g. solar power, wind power, battery storage systems etc., in a micro-grid architecture. To achieve this, advanced methods for controlling and optimally managing the heterogeneous sources are required to simultaneously target all three aspects of the energy trilemma: sustainability, affordability and security.

The ADEPT concept is based on the combination of two novel technologies:

- a novel smart inverter design that seamlessly integrates different DERs and supports the grid in providing virtual inertia with current limiting capabilities under faulty or abnormal grid conditions and
- an intelligent monitoring and control platform to manage the power flow between the DERs based on a model predictive control (MPC) framework.

## The ADEPT team

The ADEPT project involves two main partners. UK company Infinite Renewables, which leads the project, and The University of Sheffield.

The University of Sheffield team offers expertise in novel smart inverter design and the optimal control and power management system. The team is supported by four industrial partners that offer the relevant expertise for successful project delivery:

1. Yuasa Battery Europe: Battery design, integration and health monitoring
2. Swanbarton: Battery energy management system design
3. HiT Power: Inverter design
4. Heatherose: Hub and monitoring system design

## 2 Identify targeted areas of improvement against the current industry practice

The analysis identified six areas for improvement:

1. **Smart inverter design**: The existing inverter technology causes a reduction to the grid inertia, fails to support the grid under faulty conditions and requires switching between different control strategies in order to maintain a limited current from the energy source that can lead to an unpredicted and unstable behaviour. The ADEPT smart inverter aims to achieve all previous tasks in a novel unified control framework suitable for any DER.

2. **Distributed Control and Monitoring**: The optimal power management of the micro-grid will be achieved by means of distributed control and optimisation mechanisms that do not require full communication links between all DERs (centralised control). Each DER will communicate with its neighbour, paving the way for an easy 'scaling up' to multiple ADEPT units in the future.

3. **System model and state estimation**: The optimal system operation requires good knowledge of the system dynamics and states. The ADEPT solution will incorporate state estimation mechanism and prediction tools to acquire information on system states, e.g. battery state of charge, predicted wind energy.

4. **Multi-Energy Systems integration**: ADEPT aims to combine different renewable energy resources, including a 500kW wind turbine, a 100kW battery storage system and a 3kW photovoltaic system next to a Yuasa Battery site that represents the local consumer.

5. **Energy Storage and Optimal Use**: A unique combination of lead-acid and li-ion batteries without separate power conversion units will be used to introduce advanced energy storage characteristics that will be used through an optimal power management algorithm.

6. **Live Monitoring and Dashboard**: Live data from the energy resources and load demand will be collected and monitored via a dedicated monitoring system both on-site and remotely. The aim is to additionally enable the control of the power management system and adjustment of optimal control algorithm remotely from the University of Sheffield laboratory. [71]

## 3 Identify proposed benefit generated by SLES against the industry norm

These improvements are designed to have the following effects:

- **Multiple Source Optimisation System**: A systematic framework for integrating and optimally operating multiple DERs will be generated.

- **Current-limiting Smart Inverter**: A complete smart inverter device equipped with novel current-limiting control for integrating different DER units to the grid will be introduced to the industry.

- **Remote monitoring and control in (near) real-time**: An online monitoring and control platform that collects instantaneous values of power generated, stored, demanded etc., and enables flexible adjustment of the supervisory control, will be developed.

- **Manufacturing Readiness level (MRL) improvement of employed technologies**: The dual-chemistry battery will be manufactured and integrated to the grid for the first time at the power level of 100kW.

## 4 Produce a system architecture overview

The overall architecture of the ADEPT system is designed with smart elements in mind. Attaching each of the DER elements to Schneider Electric Smart Meters transforms them into smart objects that have interfaces to other systems.

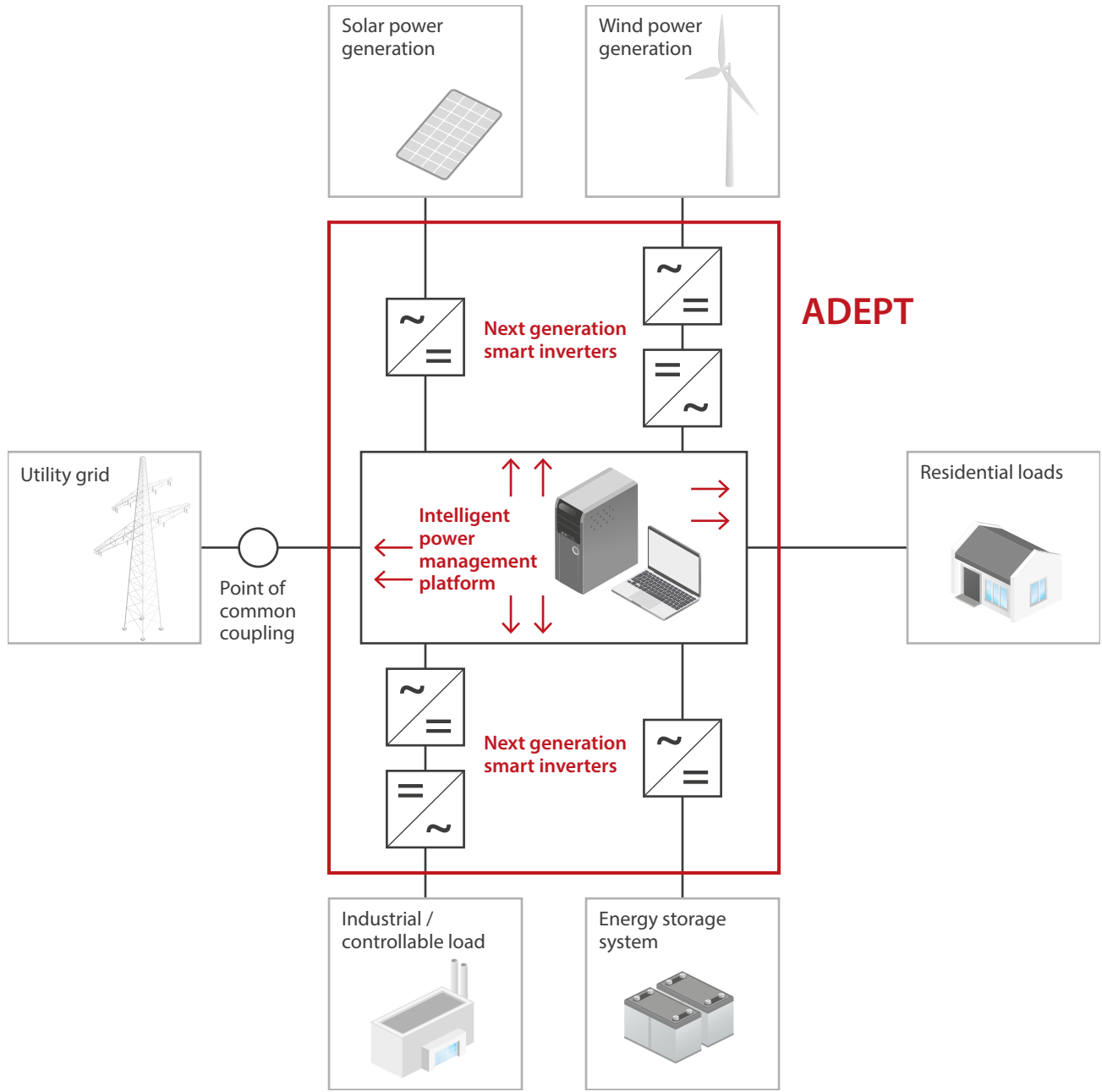

Figure 2: High Level Architecture of ADEPT.

The system architecture, as seen in Fig 2, integrates each DER unit into the ADEPT micro-grid via novel smart inverter technology that provides smooth synchronisation, grid support and reliable operation under grid variations/faults. At the centre of each element is the Hub that gathers all of the information, displays it on a dashboard and interfaces with other devices. This Hub then connects with their Multi System Controller, which is implemented on an OPAL-RT real-time digital simulator.

The residential load unit in Fig 2 represents the Yuasa Battery site. A Smart meter equipped with a GSM antenna is attached to it. This Smart meter connects to the Hub through a Modbus connection and send updates on power consumption. This information is collected to obtain a demand profile which is used as a model prediction in the optimisation algorithm.

The Wind turbine is located 0.2miles away from the ADEPT cabinet and is equipped with a dedicated commercial power conversion circuit. This represents an uncontrollable renewable source since it is designed to operate always at the point of maximum power injection to the grid. The power generation data is transmitted through secure Wi-Fi connection to the Hub in the ADEPT cabinet and is used to predict wind generation, similar to the load case, for the optimal power management algorithm.

The solar panels and the battery bank represent the two main controllable units in the ADEPT demo. They are both integrated to the grid via a novel smart inverter design that enables seamless integration and control of power injected or drawn from the main grid.

The battery bank has its own monitoring and proprietary communication system with the Hub that is based on Modbus. This system allows the user to read the battery state of charge values and to set the power output through the Smart inverter unit. The instantaneous power of the battery is transmitted to the Hub through a local CAN network. For the solar power generation system, a Smart meter is installed to directly transfer the actual instantaneous injected power to the Hub.

The supervisory control system sits on top of the individual units and connects to the Hub. It receives data from every unit and defines the desired power set points for the DER units. It is based on an OPAL-RT device onto which the distributed multi energy control system is deployed using a distributed MPC framework. The visualisation is based on a SCADA-based DAQ Factory and shows a live view of the system with instantaneous measurements of the power generated or consumed by each DER.

## Developing the system architecture

The original design of the power management and optimisation system was further expanded by introducing a transmitter at the Hub located in the ADEPT cabinet and a receiver at a remote location (University of Sheffield) through secure 3G connection. All the power generation, consumption and state of charge data) is transferred to the remote location where the OPAL-RT unit is placed. The output of the optimisation algorithm is transmitted back to the ADEPT, setting the desired power reference values to the units. This enhances system flexibility and maintenance, since supervisory control can be easily reconfigured and improved without requiring physical access to the ADEPT cabinet.

# 5 Identify the Smart properties

## Flexibility

The Smart-meters and Modbus based interconnectivity of the system mean that adding new devices and components is comparatively easy. With the Gateway-based systems the CAN bus can be extended to increase the functionality of the local system. The OPAL-RT controller deployment system allows the models to be suitably altered either on-site or from a remote location.

### Scalability

Due to the Smart device-based architecture, the edge scalability of the system is endless, provided enough IP addresses can be allocated and the network providers can cope with high demand. The bottleneck of the system is at the Hub, which is required to connect the control with the edge devices. However, the distributed nature of the supervisor control enables the integration of multiple ADEPT units in order to scale up the system with minimum reconfiguration.

### Interoperability

Any other supervisory control units using the same protocols as OPAL-RT can talk to the standard CAN and Modbus connections because these are industry standard protocols. When it comes to data analytics and big data systems relatively few systems provide support for Modbus, requiring brokers to be put in place between the two protocols.

### Prediction, intelligent control and management

The simulation and state estimation are based on advanced models running on historic data. These models are deployed in real-time systems in a hardware in the loop scenario to test the Distributed Control systems.

### Resilience and fault tolerance

The control and smart inverter technology deployed on ADEPT is comprised of self-contained systems connected through RS232 and CAN. This allows parts of the system to fail while others maintain functionality. Local monitoring and logging elements allow resilience in tracking the unit.

### Privacy and security

The mobile Internet-based connection allows the devices to have unique IP-s on the web. This makes them easier targets but reduces the chances of man in-the-middle attacks, reducing these to the service provider level, or physical snooping of signals. The use of Modbus in such a way that limits the control capability of entities reduces the risk of these. Currently all control systems run as a secondary loop on top of a primary safety one that ensures no harmful signals can be transmitted.

## 6 Review the technical and technological requirements of implementing the SLES design

The demonstrator relies heavily on Industry specific technologies and protocols such as CAN and Modbus and uses production ready Smart Devices to connect their Smart Inverter and control systems.

The supervisory control system runs on the OPAL-RT device. OPAL-RT requires its dedicated software for programming and debugging; however, it offers multiple options for communicating with the Hub. The Hub is a PLC and GPRS Module based system.

The Smart meters from Schneider that are connected to the local devices and their capability to communicate in different protocols is crucial to the systems interconnectivity.

# 7 Develop a KPI Map

The KPI Map in Fig 3 of the ADEPT Demonstrator highlights how the different components fit together and how they relate to higher level road-map.

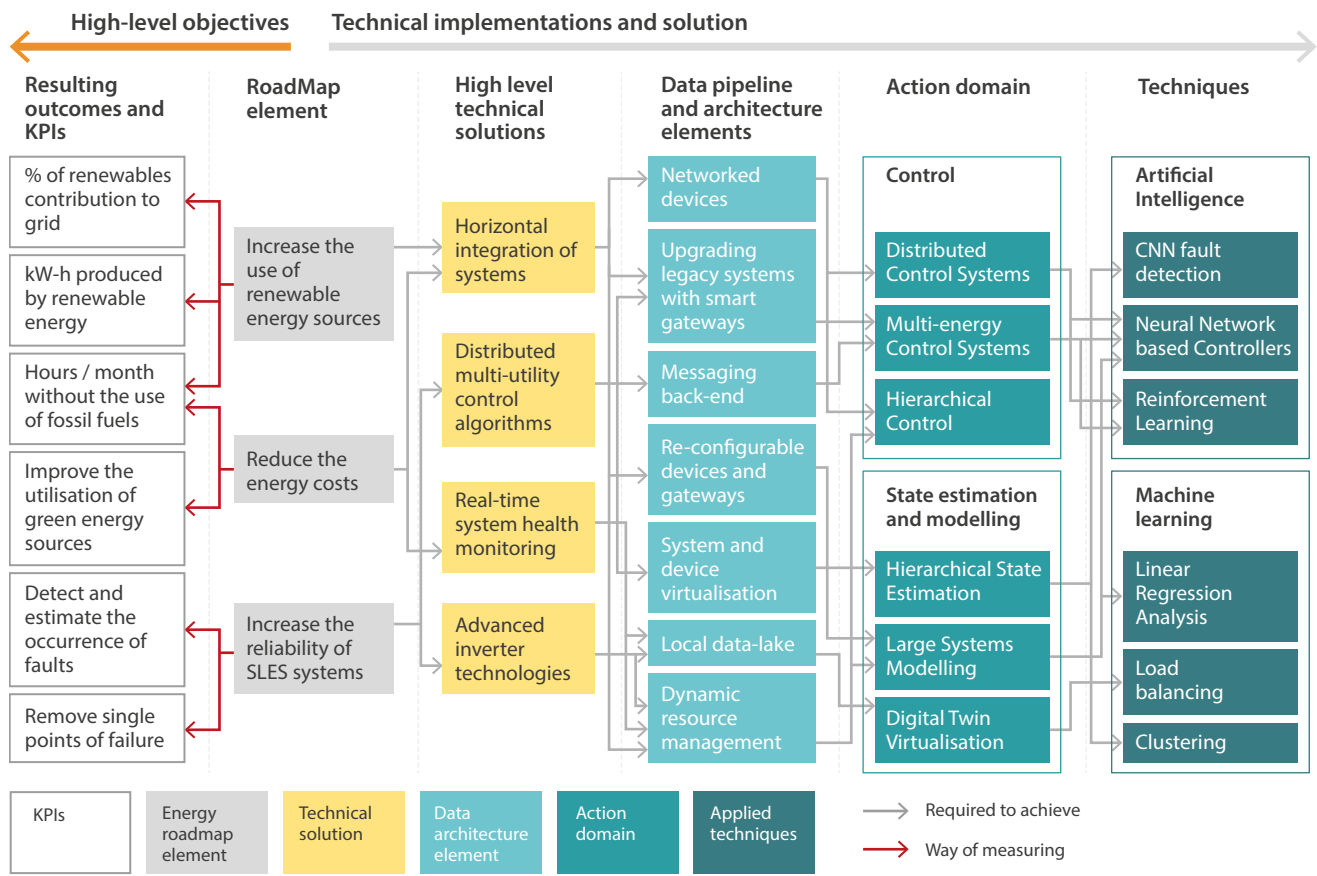

Figure 3: KPI Map of ADEPT.

# Putting the framework to work: the ADEPT Demonstrator Expansion

## 1 Determine the future technical and technological requirements

A natural extension to this project is to include more interconnected units (Batteries, Photovoltaic Panels, Wind Turbines and Energy Loads). To this aim, ADEPT like systems need to be tested in various environments before deployment to prove their robustness and demonstrate their capability. To enable this emulation of different renewable sources, storage and consumers is needed. It also requires a negotiation framework capable to update existing optimisation criterion based on current prices and market behaviour. These elements would facilitate the scaling of ADEPT in terms of number of units involved in the network and the power rating. The system also needs to be deployed among other peers, so it requires a flexible and readily available communication systems that allows various systems and services to connect to it.

## 2 Develop the new system architecture

The expansion of the ADEPT System has at its core the idea that, with added digitalisation and plug and play capabilities, testing and developing more advanced systems becomes faster and so does the benchmarking of technologies. This helps accelerate the technologies up the TRL ladder towards ruggedizing the techniques for wider deployment.

The proposed extension can be seen in Fig 4 where the different components, their links and messaging architecture is detailed.

This can be achieved through changes in the following areas.

### Data Pipelines and Architectures

- Implementing a Messaging-based Asynchronous Publish and Subscribe backend for data movement allows multiple sinks and sources to be added and removed, as well as intermediate processing to take place.

- Deploying a local data-lake that can house the existing data as well as results from tests and simulations allows for the better management of the data life-cycle and reduces the risk of losing data and adds capabilities for easier live-analysis/monitoring.

- Dynamic Resource management allows virtual devices and physical ones to interact and to be connected and organised. A system like this can help set up, optimise and deploy SLES and Local Grid structures.

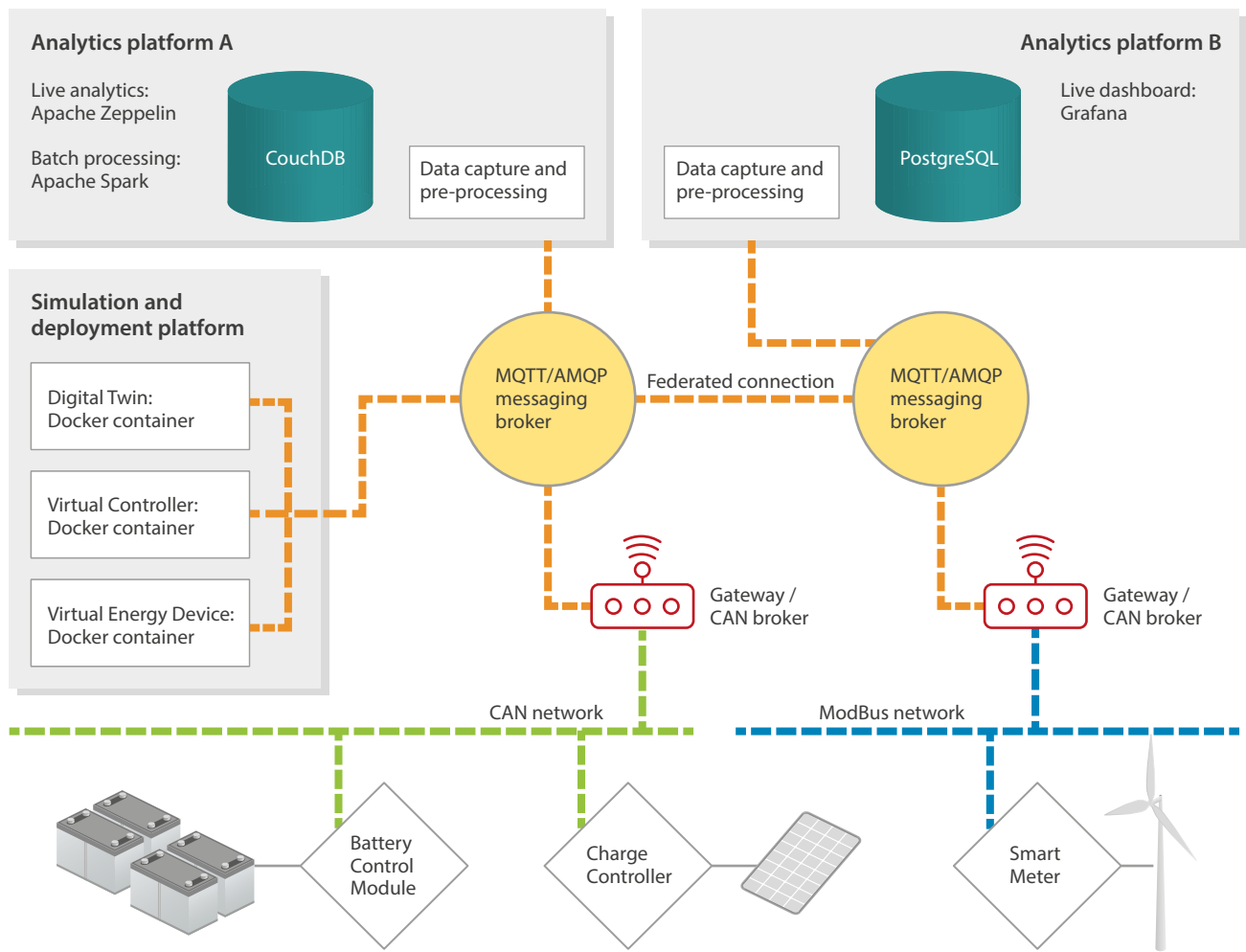

Figure 4: Extension proposal for ADEPT.

## Control Systems

- Implementing and deploying distributed multi-energy control systems.
- Deploying reconfigurable and reprogrammable Hardware in the Loop systems on the fly.
- Provide scaling analysis of algorithms for large scale, small, medium and regional energy systems.
- Extending the MPC-based microgrid control system with agents that take over the communication between peers and provide continuous cost information to inform MPC calculations and control decisions.

## State Estimation and Modelling

- Provide a Hierarchical State estimation and modelling environment where the interaction between SLES and Country wide grids can be examined.
- Use of environmental data and usage data on a larger scale can prove a more reliable testing ground for control elements.
- Digital Twin and Virtualisation to model and virtualise physical systems closely.

### Advanced Methods in CPS

- Machine learning methods can be used in state estimation problems and load balancing of systems for the architecture
- Neural Network and similar methods can be used in more advanced control scenarios
- Advanced optimisation techniques can be used to cluster and organise SLES systems.
- Reinforcement learning can be deployed at the centre of such systems to explore the capabilities and limitations of the technology and how it fares compared to traditional methods.

## 3 Evaluate the extended architecture and functionality improvements

The deployment and integration of the control, negotiation and analysis components required the creation of data brokers between their local protocols and the messaging architecture. These brokers are usually bi-directional where they translate messages to and from the AMQP messaging system which the passes them to the right location. Tailored brokers were implemented to: pass digital messages from the agents' framework through their XMPP messaging system to AMQP; pass messages from the real-time and MATLAB based controllers using UDP ports; to store logging and monitoring data in the data stores and to retrieve the requested information for the agents and MPC controllers.

These brokers add extra latencies to the system but ensure that the components can function in the environment they are designed for, making use of all the existing functionality for reporting and evaluation.

# Conclusions

It is a time consuming and intensive process to gather the information required to undertake the steps in this detailed analysis framework. The information is often not readily available and generally requires personal contact with the original project team; moreover, information erodes over time. However, the effort required will be worth it because it breaks down the SLES demonstrator into its core cyber-physical components, allowing researchers and industry alike to evaluate the best next development steps and approaches along with the benefits these could generate. It also provides a framework which ensures future flexibility and scalability is built into the energy system's cyber-physical functions.

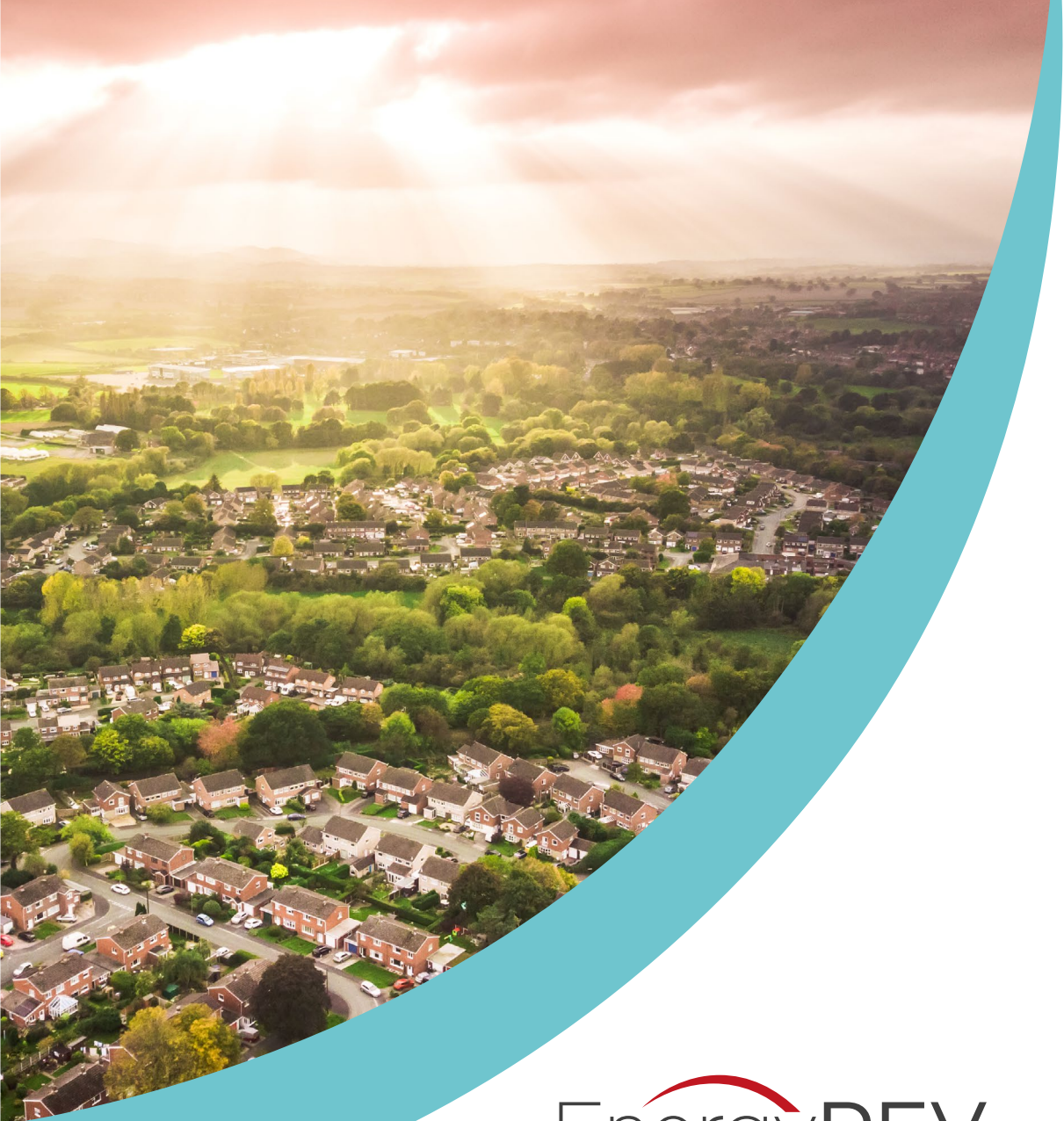

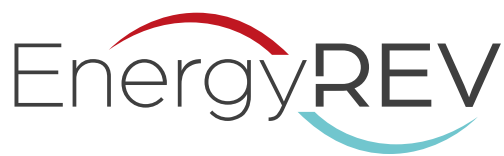

## Want to know more?

www.energyrev.org.uk

@EnergyREV_UK

info@energyrev.org.uk

Sign up to receive our newsletter and keep up to date with our research, or get in touch directly by emailing info@energyrev.org.uk

### About EnergyREV

EnergyREV was established in 2018 (December) under the UK's Industrial Strategy Challenge Fund Prospering from the Energy Revolution programme. It brings together a team of over 50 people across 22 UK universities to help drive forward research and innovation in Smart Local Energy Systems.

ISBN 978-1-909522-58-9

EnergyREV is funded by UK Research and Innovation, grant number EP/S031863/1

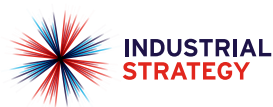
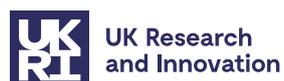